\documentclass[fleqn,usenatbib]{mnras}

\usepackage[T1]{fontenc}

\DeclareRobustCommand{\VAN}[3]{#2}
\let\VANthebibliography\thebibliography
\def\thebibliography{\DeclareRobustCommand{\VAN}[3]{##3}\VANthebibliography}

\usepackage{graphicx}	
\usepackage{amsmath}	
\usepackage{amssymb}	

\title[Pseudo labelling for galaxy properties]{Improving machine learning-derived photometric redshifts and physical property estimates using unlabelled observations}

\author[A. Humphrey et al.]{
Humphrey, A.,$^{1,2}$\thanks{E-mail: andrew.humphrey@astro.up.pt (AH)}
Cunha, P.~A.~C.$^{1,3}$
Paulino-Afonso, A.,$^{1}$
Amarantidis, S.,$^{4}$
Carvajal, R.,$^{5,6}$
\newauthor
~Gomes, J.~M.,$^{1}$
Matute, I.,$^{5,6}$
Papaderos, P.$^{5,6}$
\\
$^{1}$ Instituto de Astrof\'{i}sica e Ci\^encias do Espa\c{c}o, Universidade do Porto, CAUP, Rua das Estrelas, Porto, 4150-762, Portugal\\
$^{2}$ DTx -- Digital Transformation CoLAB, Building 1, Azur\'em Campus, University of Minho, 4800-058 Guimar\~aes, Portugal\\
$^{3}$ Faculdade de Ci\^{e}ncias da Universidade do Porto, Rua do Campo de Alegre, 4150-007 Porto, Portugal\\
$^{4}$ Institut de Radioastronomie Millim\'{e}trique (IRAM), Avenida Divina Pastora 7, Local 20, E-18012, Granada, Spain\\
$^{5}$ Departamento de F\'isica, Faculdade de Ci\^encias, Universidade de Lisboa, Edif\'icio C8, Campo Grande, PT1749-016 Lisboa, Portugal\\
$^{6}$ Instituto de Astrof\'isica e Ci\^encias do Espa\c{c}o, Faculdade de Ci\^encias, Universidade de Lisboa, Tapada da Ajuda, PT-1349-018 Lisboa, \\ ~~Portugal\\
}

\date{Accepted 5 December 2022. Received 15 November 2022; in original form 9 June 2022}

\pubyear{2022}

\begin{document}
\label{firstpage}
\pagerange{\pageref{firstpage}--\pageref{lastpage}}
\maketitle

\begin{abstract}
In the era of huge astronomical surveys, machine learning offers promising solutions for the efficient estimation of galaxy properties. 
The traditional, `supervised' paradigm for the application of machine learning involves training a model on labelled data, and using this model to 
predict the labels of previously unlabelled data. The semi-supervised `pseudo-labelling' technique offers an alternative paradigm, allowing 
the model training algorithm to learn from both labelled data and as-yet unlabelled data. We test the pseudo-labelling method on the problems 
of estimating redshift, stellar mass, and star formation rate, using COSMOS2015 broad band photometry and one of several publicly available machine 
learning algorithms, and we obtain significant improvements compared to purely supervised learning. We find that the gradient-boosting tree methods 
\texttt{CatBoost}, \texttt{XGBoost}, and \texttt{LightGBM} benefit the most, with reductions of up to $\sim$15\,\% in metrics of absolute error. 
We also find similar improvements in the photometric redshift catastrophic outlier fraction. We argue that the pseudo-labellng technique will be useful 
for the estimation of redshift and physical properties of galaxies in upcoming large imaging surveys such as {\it Euclid} and LSST, which will provide 
photometric data for billions of sources. 
\end{abstract}

\begin{keywords}
methods: statistical -- galaxies: photometry -- galaxies: fundamental parameters -- galaxies: distances and redshifts
\end{keywords}



\section{Introduction}
The Universe and its evolution cannot be understood without first understanding galaxies, and their evolution.
A key endeavour in extragalactic astrophysics is the estimation of the redshift and physical properties of galaxies (e.g., stellar mass, star-formation 
rate, etc.), thereby allowing galaxies to be placed in evolutionary sequences and frameworks, and facilitating the testing of 
galaxy evolution models \citep[e.g.,][]{Forster2020}. 

Galaxy property estimation methods have traditionally used software that fits spectral templates to observed spectral energy distributions 
or spectra \citep[e.g.,][]{Arnouts1999,Bolzonella2000,CidFernandes2005,Ilbert2006,daCunha2008,Noll2009,Laigle2016,Gomes2017,Carnall2018,Johnson2021}; 
while generally effective and physically motivated, template fitting tends to be expensive and often does not scale well to very large 
datasets (i.e., $\gg1{\rm M}$ objects), since the computation time typically scales linearly with the number of sources to be fitted. 

Supervised machine learning methods represent an alternative that is increasingly applied to the problems of 
galaxy classification and property estimation. In this paradigm, a learning algorithm infers a function to map observations
to labels, using examples of observation and label pairs. The function is subsequently used to predict the labels of new (unlabelled) 
examples \citep[see, e.g.,][]{Baron2019}. A major benefit of such methods is their usually greater scalablity compared to template fitting 
methods: Once trained, machine learning models can usually be applied to large volumes of unlabelled data at a negligible computational 
cost \citep[e.g.,][]{Hemmati2019}. 

There are now numerous studies using supervised machine learning in the field of extragalactic astrophysics. Among its most 
frequent uses is in the separation of sources into different classes, for which there is now a rich body of literature. 
For instance, supervised learning has been successfully used by various authors to perform binary classification with the goal 
of selecting sources with a specific set of characteristics \citep[e.g.,][]{Cavuoti2014}, or for multi-class problems 
such as star/galaxy/quasar classification \citep[e.g.,][]{Bai2019,clarke2020,Cunha2022}

Supervised machine learning has also been extensively used to estimate galaxy redshifts from imaging or photometric data, via 
regression \citep[e.g.,][]{Collister2004,Brescia2013,Cavuoti2017,Pasquet2019,Razim2021,Guarneri2021,Carvajal2021,Cunha2022}. While
reducing the required computational time is often among the main motivations for using machine learning to estimate redshift, 
there are now indications that these methods can significantly outperform traditional template-fitting methods, in some regions of 
colour-redshift space, in terms of the accuracy estimated redshifts \citep[][]{Desprez2020}.
 
Other key uses of supervised learning in extragalactic astronomy are the estimation galaxy physical properties 
\citep[e.g.,][]{Bonjean2019,Delli2019,Mucesh2021,Simet2021}, and morphological classification of galaxies from images 
\citep[e.g.,][]{Dieleman2015,Huertas-Company2015,DominguezSanchez2018,Tuccillo2018,Nolte2019,Bowles2021,Bretonniere2021}. 
A few studies have also explored schemes for combining results from traditional methods with machine learning methods, 
resulting in improved predictions \citep[e.g.,][]{Cavuoti2017,Fotopoulou2018}. 

One weakness of the standard supervised learning paradigm is that only labelled data can be used for model training, and 
potentially useful information that may exist within the unlabelled data remains invisible to, and thus unused by, the 
training algorithm. As such, it is interesting to consider methodologies whereby models can be trained with an awareness
of the content of unlabelled test data. 

The `pseudo-labelling' technique is a semi-supervised machine learning method that makes use of unlabelled data to
refine supervised machine learning models \citep{Lee2013,Slijepcevic2021}. By allowing model training to make use of information 
about the structure and compostion of the unlabelled data, the resulting model can more accurately predict labels 
for this data. In this paper, we demonstrate the application of pseudo-labelling to the estimation of galaxy properties using
supervised machine learning and broad-band photometry.

We begin by giving a brief overview of the pseudo-labelling methodology in $\S$\ref{methods}. The data and its preprocessing
are described in $\S$\ref{data}. The application of pseudo-labelling to machine-learning-based estimation 
of photometric redshifts is presented in $\S$\ref{photoz}. Examples of its application to the estimation
of stellar mass and star formation rate are presented in $\S$\ref{mass_sfr}. Our conclusions and final remarks are  
given in $\S$\ref{conclusions}.

\section{The pseudo-labelling methodology}
\label{methods}
Pseudo-labelling is a semi-supervised technique which may be incorporated into supervised machine-learning pipelines,
leveraging information contained within unlabelled data such that stronger models may be produced \citep{Lee2013}.
The main improvements from this technique arise by allowing supervised learning algorithms to obtain a clearer 
picture of the structure of the overall dataset from which the training and hold-out data are drawn. This includes
allowing the learning algorithm to recognise when `outlier' examples in the training set are in fact members of a significant
population that is worthwhile to model rather than to ignore. 

In its simplest form, the pseudo-labelling workflow is as follows. First, a supervised machine learning model is trained
on a set of labelled training data. The resulting model is {then used to predict the labels for the unlabelled (test) 
data. Subsequently, a new, larger training set is created by appending the test data to the original training data. 
When creating this new training set, the examples that come from the test set are labelled with their predicted labels 
(the `pseudo-labels'). The labels associated with the original training data remain unaltered, however. 
This yields a significantly larger training set, where the labels used for training are now a mixture of 
ground-truth values, and values predicted by the aforementioned model.

A new model is trained on this new training set, and is used to predict new labels for the test set examples. 
This process may be repeated for an arbitrary number of iterations, each time updating the pseudo-labels to their most recent 
predicted values, prior to training the next model. After the desired number of iterations has been completed, the resulting 
model is used to make the final label predictions for the test set (and other hold-out sets, if applicable). Our pseudo-labelling 
workflow is illustrated in Fig. ~\ref{workflow}. 

{A real world application of this workflow might be the prediction of photometric redshifts using broad-band photometry 
from the {\it Euclid} Wide Survey \citep{Scaramella2021}.

In this study, we perform 50 iterations of pseudo-labelling before stopping. This was decided in order to have a 
reasonable balance between computation time and number of iterations. The results presented in this work required $\sim20 {\rm h}$ 
of computation time on a workstation with an Intel Core i7 11700K 8-core CPU (3.6-5.0 GHz), 32 GB of RAM, and an NVIDIA RTX A6000 48GB GPU. 
However, the code can also be executed without a GPU, or with less powerful hardware, with some minor modifications. 

We test the pseudo-labelling methodology with five open-source learning algorithms. Three of these are gradient-boosting tree (GBT) algorithms:  
\texttt{CatBoostRegressor}\footnote{\href{https://catboost.ai}{https://catboost.ai}} \citep{Prokhorenkova2018},
\texttt{LGBMRegressor}\footnote{\href{https://lightgbm.readthedocs.io}{https://lightgbm.readthedocs.io}} \citep{Ke2017},
and \texttt{XGBoost}\footnote{\href{https://xgboost.readthedocs.io}{https://xgboost.readthedocs.io}} \citep{Chen2016}. 
{While there are subtle differences between them, the GBT methods all have a fundamentally similar approach to model
 training: they combine multiple regression trees in series, with each new tree fitting the residuals from the previous 
 trees \citep[`boosting'; see][]{Friedman2001}.

From the \texttt{Python} 
\texttt{Scikit-Learn}\footnote{\href{https://scikit-learn.org}{https://scikit-learn.org}} package \citep{Pedregosa2011}
we also use the popular tree-ensemble method \texttt{RandomForestRegressor} \citep{Breiman2001}. 
This learning algorithm combines multiple weak regression trees into a single, stronger regression model by 
averaging the predictions from all the individual trees. To reduce the similarity between the constituent trees, each tree is built using a subset of the features and training examples that is selected at random (with replacement).

Also from \texttt{Scikit-Learn}, we test the distance-based method \texttt{KNeighborsRegressor}. Using the 
$k$-nearest neighbours (KNN) algorithm, this is a non-parametric supervised method, which uses the similarity 
(typically defined by the Euclidean distance) to training examples to predict the labels for test examples. 
For regression problems, label predictions are usually made by averaging the $k$-nearest labels from the training set, 
where $k$ is an integer specified by the user.
 
 The learning algortithms were optimized to give competitive results for the prediction problems posed herein, but 
 were not optimized exhaustively, since a test of which algorithm performs best for galaxy physical property 
 estimation is not the goal of this study. The Python scripts used in this work are publicly available at 
 Github\footnote{\href{https://github.com/humphrey-and-the-machine/pseudo-labelling}{https://github.com/humphrey-and-the-machine/pseudo-labelling}. 
The readme file for this repository gives instructions for how to modify the code for use without a GPU, and/or with less powerful computing hardware.}.

We emphasize that this methodology differs significantly from model stacking workflows \citep[e.g.,][]{Wolpert1992, Zitlau2016,Humphrey2022,Cunha2022}; whereas 
pseudo-labelling involves appending unlabelled data (with model-predicted labels) to the training data, stacking usually only makes use of 
labelled data for model training.

\begin{figure*}
	\includegraphics[clip,trim={0 100 0 0},width=2\columnwidth]{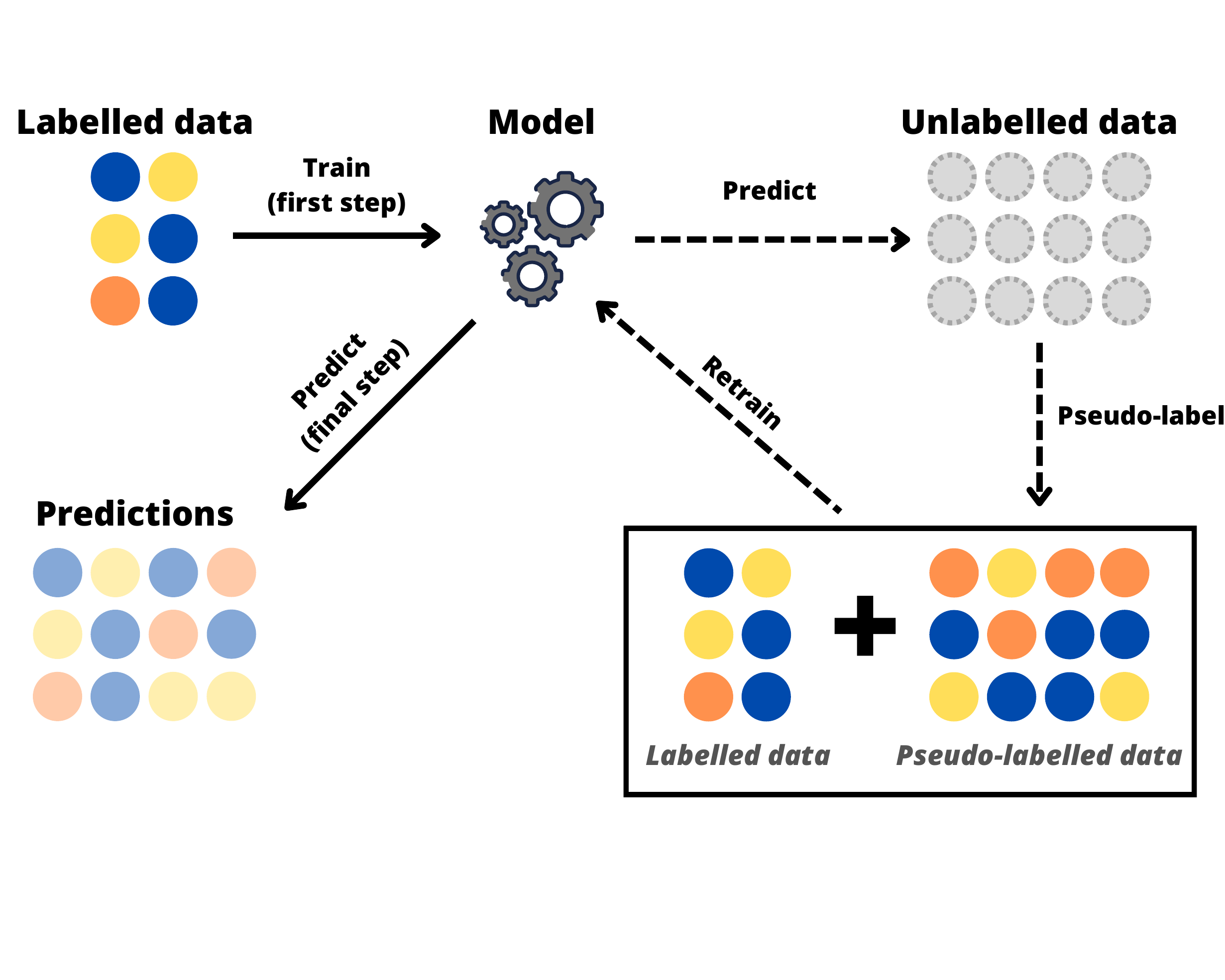}
    \caption{Flow diagram illustrating the pseudo-labelling proceedure that we have used. Labelled data is used to train an initial model;
    the model predicts (pseudo) labels for unlabelled data; this pseudo-labelled data is appended to the training data; the resulting dataset used to train
    a new model, predict new labels, update the pseudo-labels, and so on, until stopping criteria are reached (50 iterations in our case).}
   \label{workflow}
\end{figure*}

\begin{figure}
	\includegraphics[width=1\columnwidth]{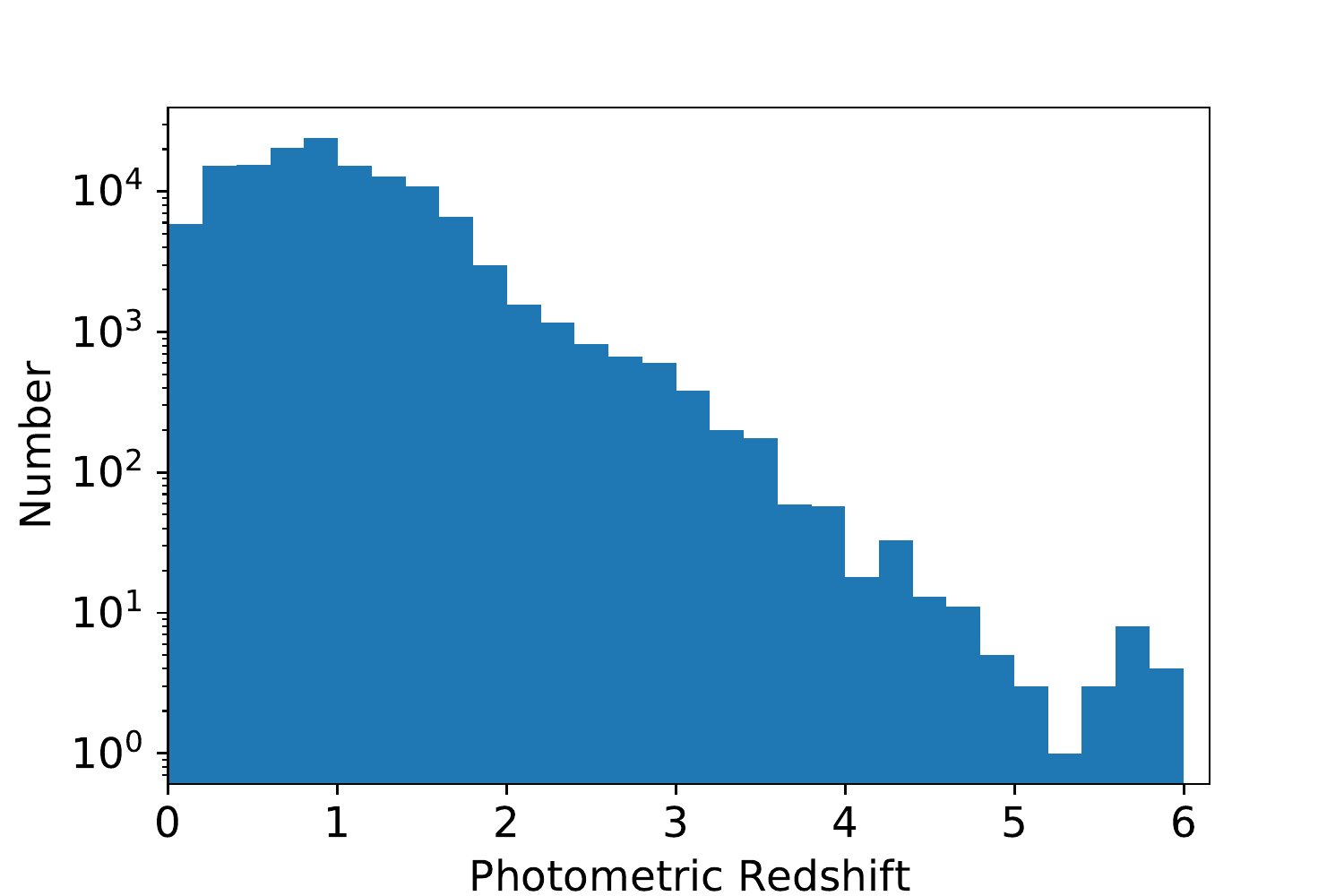}\\
	\includegraphics[width=1\columnwidth]{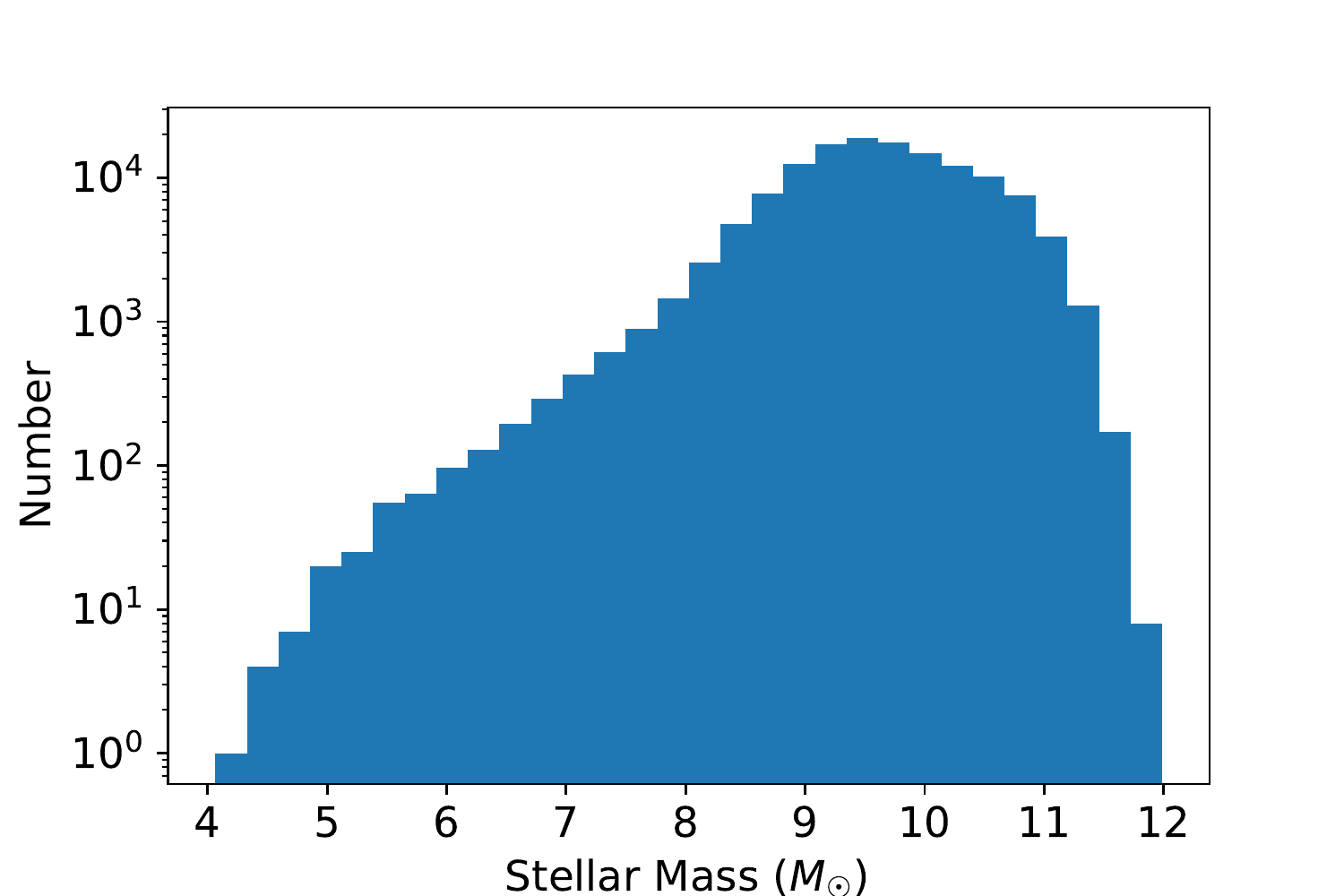}\\
	\includegraphics[width=1\columnwidth]{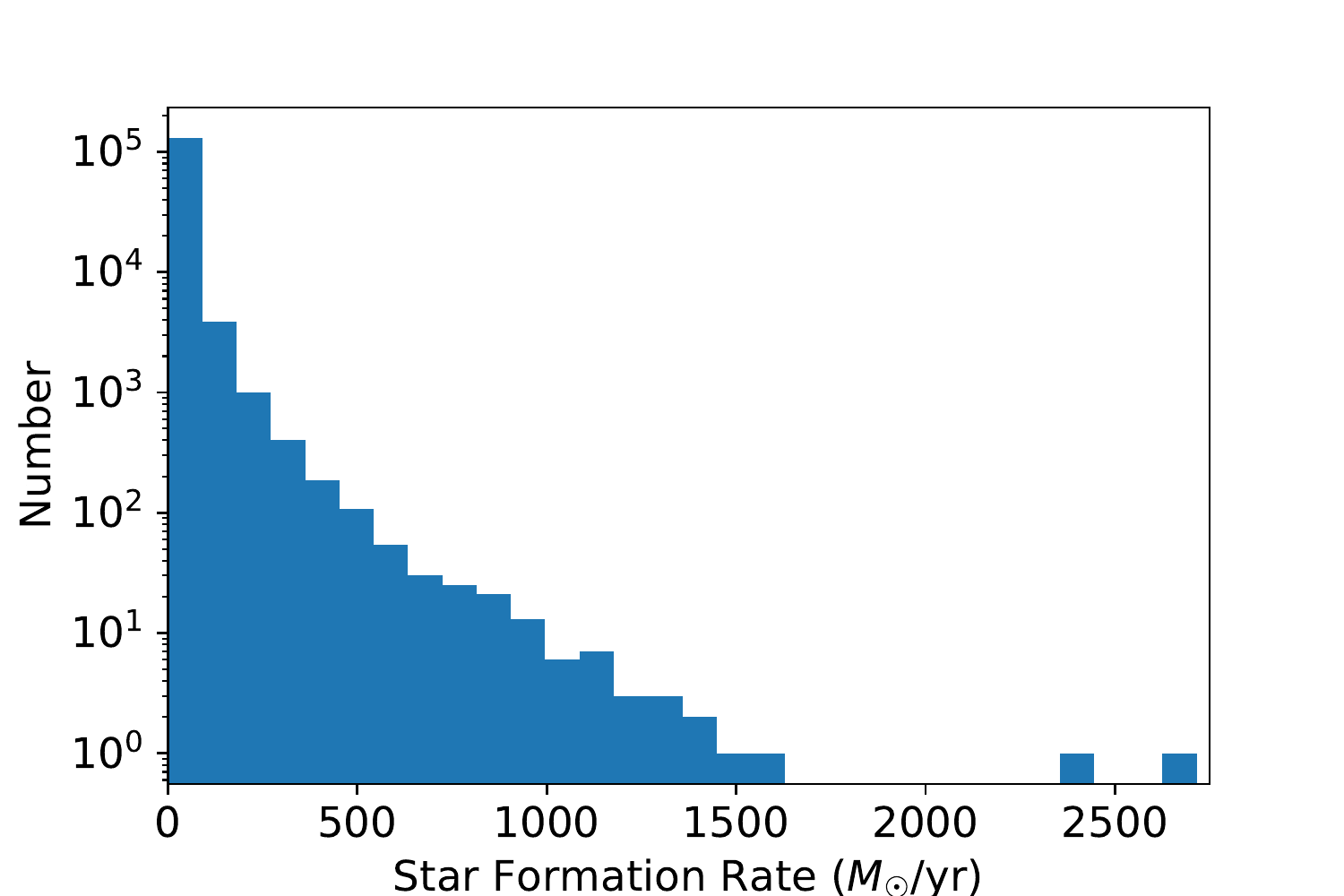}\\
    \caption{Histograms for photometric redshift, stellar mass, and star formation rate for the selected sample.}
   \label{fig:histograms}
\end{figure}

\begin{figure*}
	\includegraphics[width=2\columnwidth]{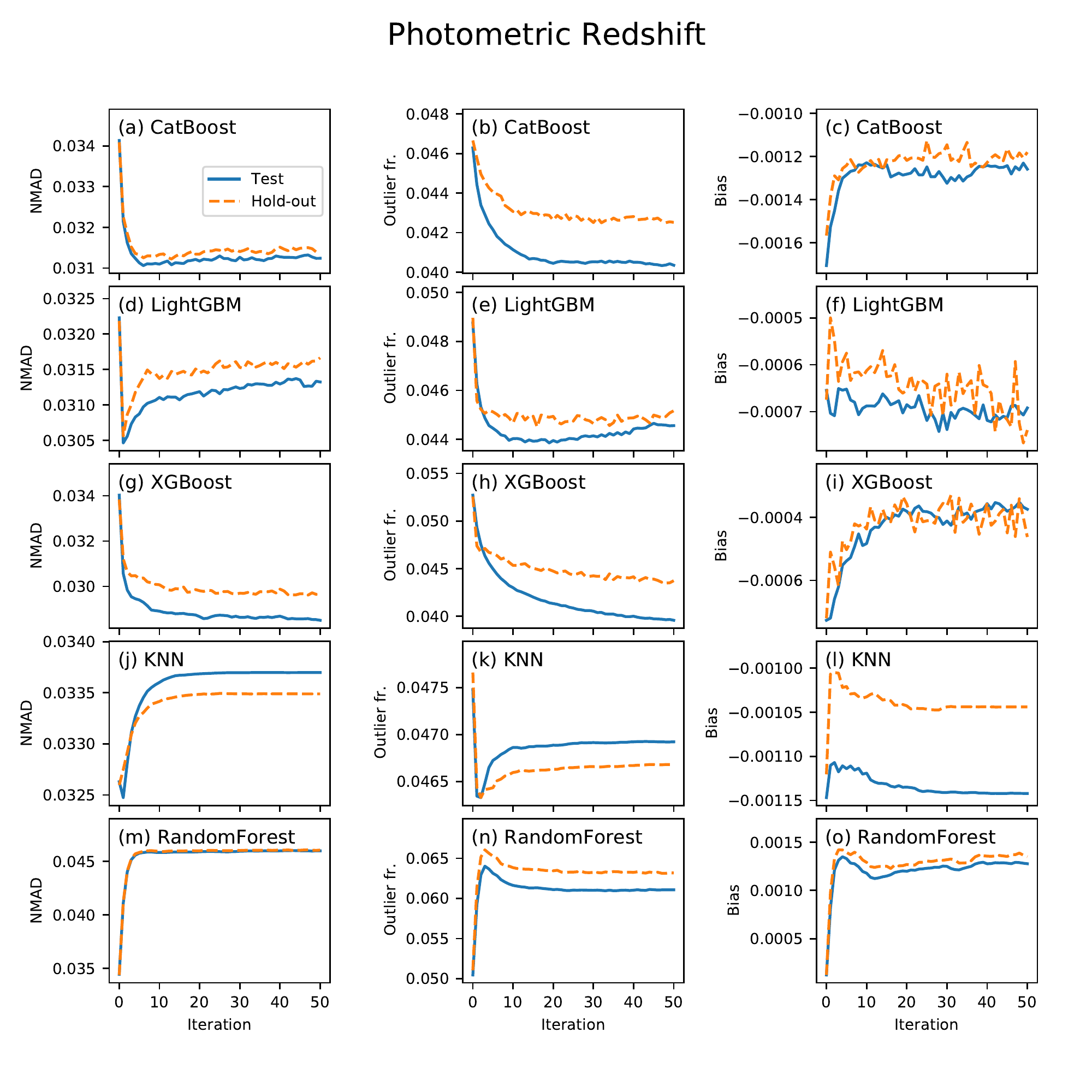}
    \caption{Estimating photometric redshifts using pseudo labelling with \texttt{CatBoost}, \texttt{LightGBM}, \texttt{XGBoost}, 
    \texttt{KNeighborsRegressor} or \texttt{RandomForestRegressor}. 
    The zeroth iteration corresponds to the initial model trained only on the training set; the subsequent 50 iterations correspond to the pseudo-labelling
    process. The solid blue line shows the evolution of the photometric redshift metrics as a function of the pseudo labelling iteration, for the test set
    used for pseudo-labelling. The dashed orange shows the evolution of the metrics for the hold-out set, which was not used for model training or pseudo labelling.
    The results shown in this figure correspond to the averaged metrics from 10 runs with different random seeds for the random train / test / hold-out split, 
    and where applicable, for the learning algorithms also.  
    Left column: Normalised median absolute deviation. 
    Middle column: Catastrophic outlier fraction. 
    Right column: The overall bias of the prediction errors.}
   \label{fig1}
\end{figure*}

\begin{figure*}
	\includegraphics[width=2\columnwidth]{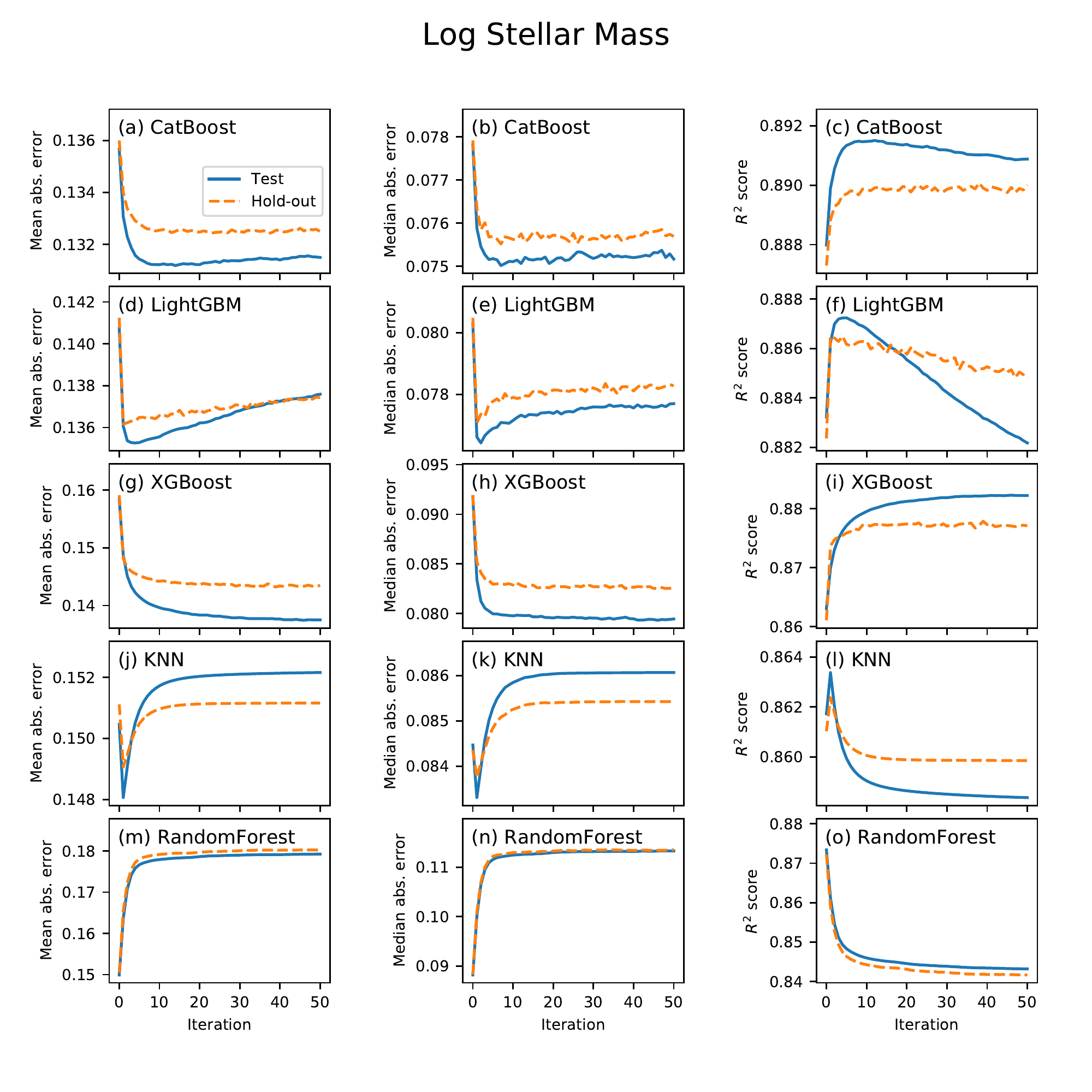}
    \caption{Similar to Fig.~\ref{fig1}, but for the estimation of $Log_{10} M_{\star}$. 
    Left column: Mean absolute error. 
    Middle column: Median absolute error. 
    Right column: $R^2$ score. }
   \label{fig2}
\end{figure*}

\begin{figure*}
	\includegraphics[width=2\columnwidth]{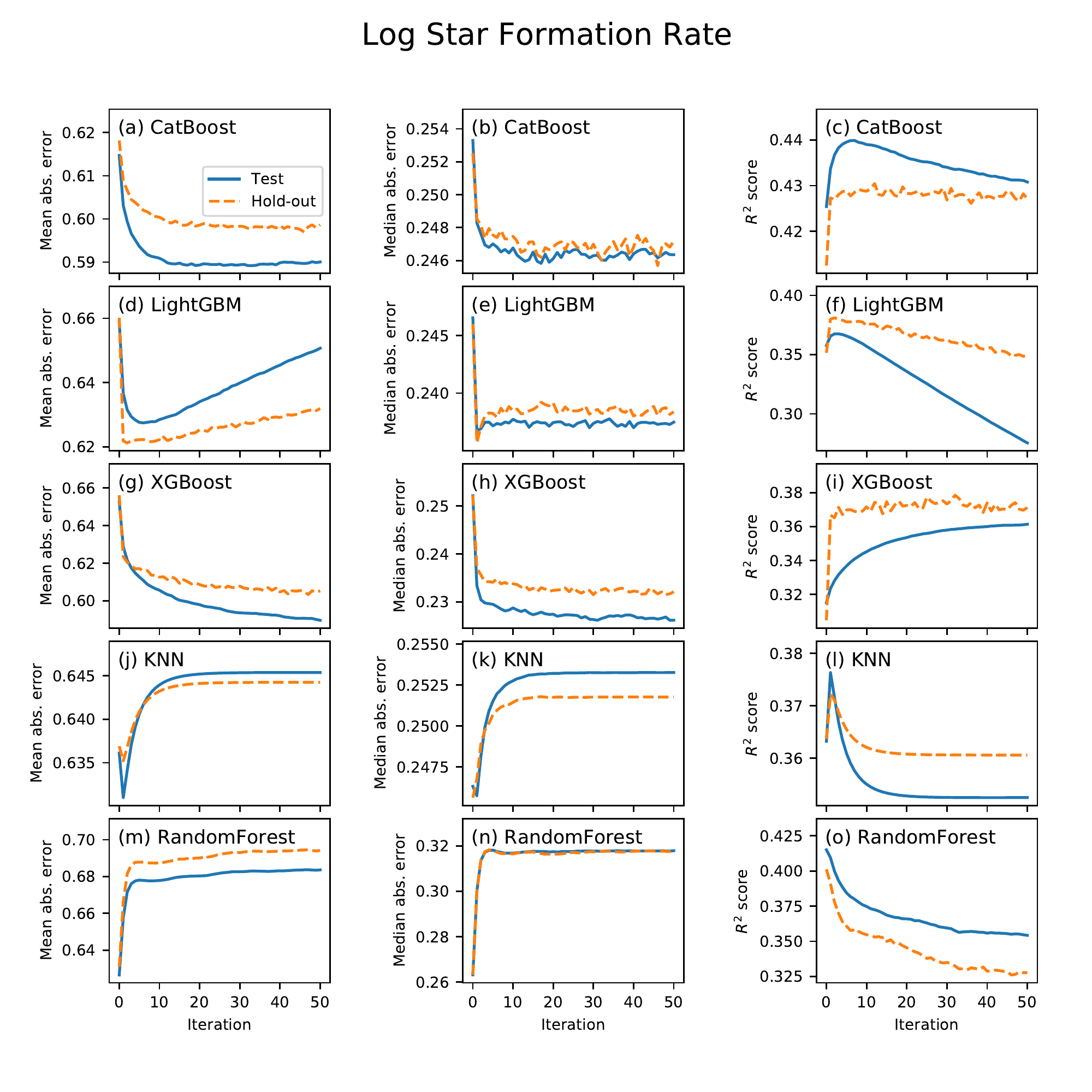}
    \caption{Similar to Figs.~\ref{fig1} and ~\ref{fig2}, but for the estimation of $Log_{10} {\rm SFR}$. }
   \label{fig3}
\end{figure*}

\begin{figure*}
	\includegraphics[width=2\columnwidth]{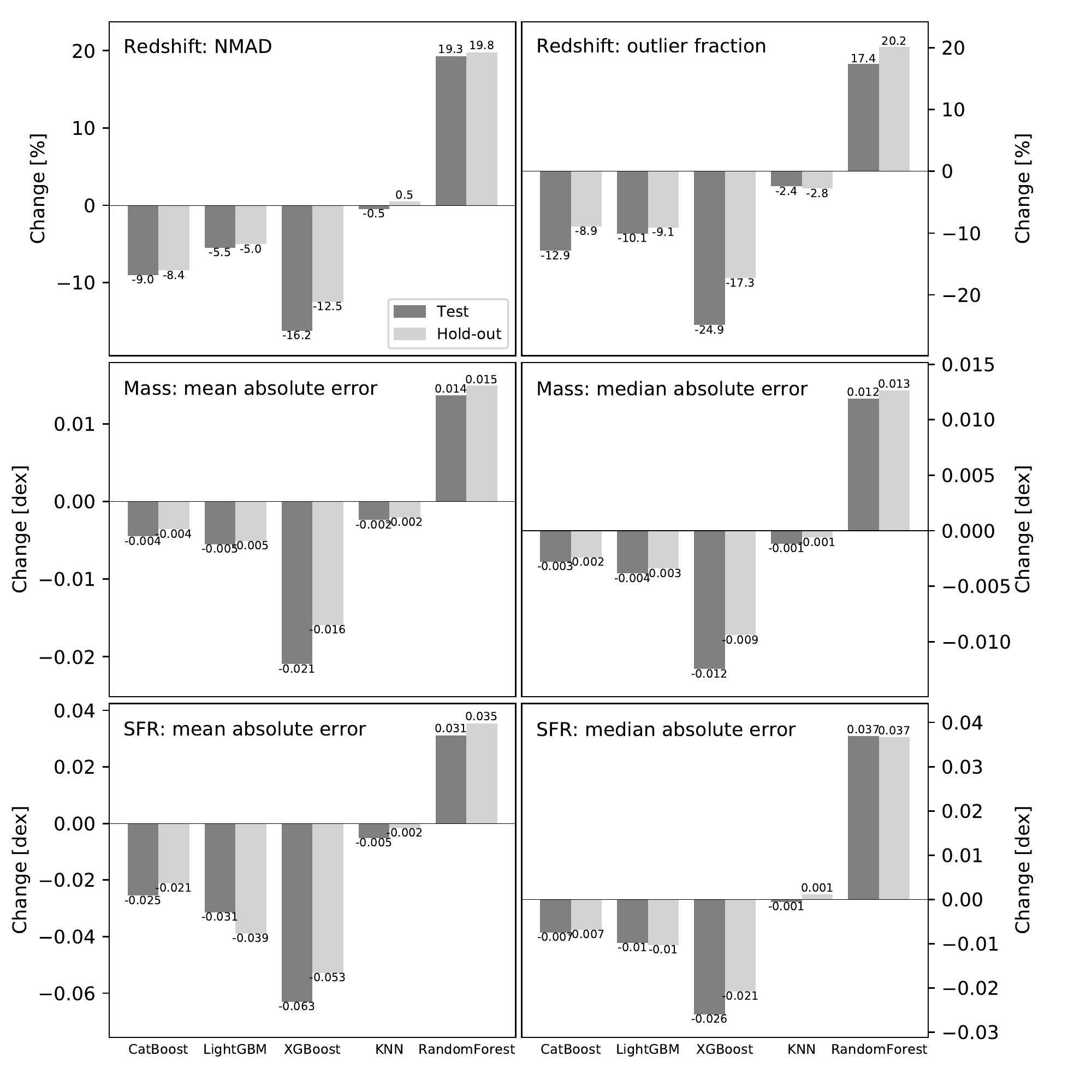}
    \caption{Quantifying the improvement, if any, obtained by applying pseudo-labelling. Each panel shows the maximum improvement obtained, 
    relative to the zeroth iteration, for each learning algorithm, during the 50 iterations of pseudo-labelling. Negative values 
    (i.e., reduced errors) indicate an improvement, and positive values indicate that model quality worsened.}
   \label{fig4}
\end{figure*}

\section{Data and preprocessing}
\label{data}
The dataset used in this work is derived from the COSMOS2015 multiwavelength catalogue \citep{Laigle2016}. We selected a subset of 
$135,374$ galaxies from the catalogue using the following criteria:
\begin{itemize}
\item (i) Galaxies not detected in X-rays (\texttt{TYPE=0});
\item (ii) photometric redshifts in the range $0 < z < 9.8$;
\item (iii) $Y$, $J$, and $H$ $\le24$ AB mag.
\end{itemize}

These criteria were designed to (i) remove active galaxies where possible, (ii) remove sources with a problematic photometric redshift, 
and (iii) remove faint sources with low signal to noise photometry. 

To create the features for model training, we extracted the $u$, $B$, $V$, $r$, $i_+$, $z_+$, $Y$, $J$, $H$ and $K_s$ broad band magnitudes, 
as measured in the 3\arcsec aperture. In the interest of simplicity and repeatability, we have used the COSMOS2015 \texttt{MAG\_APER3} measurements 
without applying any transformations or corrections. 
All unique colour combinations were then computed, resulting in 55 features (10 magnitudes and 55 colours).

Missing values represent 0.03\% of these data. At the time of writing, there is no established best-practise for imputing missing photometry 
for use with machine learning; many of the traditional imputation methods used by machine learning practioners (e.g., adopting the mean or median value) 
have the potential to result in unphysical spectral energy distributions, particularly when a photometry value is missing due to the presence of a 
spectral break in the spectrum of the source. We opted to impute missing photometry values with the maximum measured magnitude of the other galaxies 
in the same band \citep[e.g.,][]{Carvajal2021,Curran2022}; while this does not inform the learning algorithms that the magnitude value is a special case, 
it does at least place sources that have missing values at approximately correct locations within the colour-magnitude hypercube  
\citep[cf.][who imputed `magic values' to allow learning algorithms to treat missing values differently when beneficial]{Humphrey2022}.

We have adopted the COSMOS2015 values of \texttt{PHOTOZ}, \texttt{MASS\_BEST}, and \texttt{SFR\_BEST} as the ground truth values of 
the redshift, $log_{10} (M_{*}/M_{\odot})$, and $log_{10} {\rm (SFR)}$, respectively\footnote{In the interest of brevity, we hereinafter use
$M_{*}$ and SFR to mean the log of the quantities.}. These quantities are not `ground-truth' in the conventional sense, in that they were 
originally produced by applying the \texttt{Le Phare} template-fitting code to 30 COSMOS2015 photometry bands \citep{Laigle2016}. As such, 
these quantities are subject to both random and systematic errors with respect to their `true' values. 
Figure ~\ref{fig:histograms} shows histograms for photometric redshift, stellar mass, and star formation rate for the selected sample.

The dataset was then randomly split into `train', `test', and `hold-out' sets with a 12:28:10 ratio (24\%, 56\%, and 20\%), using the 
\texttt{Scikit-Learn} function \texttt{train\_test\_split}. The train set (or training set) is used for model training, with the ground 
truth labels visible to the learning algorithm. Finally, all 55 features in the three data subsets (train, test, hold-out) were standardized 
by removing their mean and scaling to unit variance, using the mean and standard deviation calculated from the train set. 

The test set simulates unlabelled data that is to be used for pseudo-labelling; our hold-out set simulated unlabelled data that is not used for pseudo-labelling, 
but which we use to test how well models generalise to data that was not used for training or pseudo-labelling 
(note that our definition of the test and hold-out sets differs slightly from some of the literature).
For the test and hold-out sets, the ground truth labels are not visible to the learning algorithms, and are used only to evaluate metrics of model performance.

\section{Metrics of model quality}
\label{sec:metrics}
To evaluate our regression models, we use several different metrics. In the case of photometric redshift estimation, we use the normalized median absolute deviation
(NMAD) as measure of accuracy, which we calculate as

\begin{equation}
    \mbox{NMAD} = 1.48 \, \mbox{median}\,\left(\frac{\,|z_{\rm phot}-z_{\rm ref}|\,}{1+z_{\rm ref}}\right)\,,
	\label{eq:nmad}
\end{equation}

\noindent where $z_{\rm phot}$ is the redshift predicted by the model, and $z_{\rm ref}$ is the reference photometric redshift used as the `ground truth'. The 
NMAD is roughly analogous to the standard deviation: $\sigma \approx \kappa \,{\rm MAD}$, where MAD is the median absolute deviation, and $\kappa$ is a 
constant scale factor; $\kappa \approx 1.48$ for normally distributed data. An advantage of the NMAD metric is that it is less affected by outliers than is the 
classical standard deviation. 
 
It is also of interest to use a metric that quantifies the number of outliers among the redshift predictions. Thus, we adopt the following criterion to classify 
redshift estimates as catastrophic outliers:

\begin{equation}
  \frac{|z_{\rm phot}-z_{\rm ref}|}{1+z_{\rm ref}} > 0.15.
	\label{eq:f_out}
\end{equation}

In addition, we define the bias of the photometric redshift estimates as

\begin{equation}
  \mbox{bias} = \mbox{median}\left(\frac{z_{\rm phot}-z_{\rm ref}}{1+z_{\rm ref}}\right).
	\label{eq:bias}
\end{equation}

To quantify the errors in predicting $M_{*}$ and SFR, we use the mean absolute error, the median absolute error, 
and the coefficient of determination $R^2$. The mean absolute error is calculated as

\begin{equation}
	\mbox{mean absolute error} = \frac{\sum\limits_{i=1}^{n}(\hat{y_i}-y_i)}{n}
	\label{eq:mean_abs_error}
\end{equation}

\noindent where $y_i$ is the true value of the label of the $i$-th sample, $\hat{y_i}$ is the predicted value, and $n$ is the total number 
of samples. Similarly, the median absolute error is calculated as

\begin{equation}
	\mbox{median absolute error} = \mbox{median}\left(\hat{y_i}-y_i\right).
	\label{eq:median_abs_error}
\end{equation}

The $R^2$ score is calculated as

\begin{equation}
    R^2 = \frac{\sum\limits_{i=1}^{n}(\hat{y_i}-y_i)}{\sum\limits_{i=1}^{n}(y_i-\bar{y})}
	\label{eq:r2}
\end{equation}

\noindent where $\bar{y}$ is $y_i$ averaged over $n$ samples. Unlike the previous metrics, higher values of $R^2$ indicate a better model, 
up to a maximum value of 1, with no minimum value.

\section{Results and discussion}
\label{results}
In Figs.~\ref{fig1}-\ref{fig3}, we show the evolution of selected metrics of model quality versus iteration number, when applying 
pseudo-labelling to the estimation of photometric redshift (Fig.~\ref{fig1}), stellar mass (Fig.~\ref{fig2}), and SFR (Fig.~\ref{fig3}). 
Each quantity was predicted separately, using a dedicated single-output model.

The curves correspond to the metrics of model quality, averaged over 10 runs, using different random seeds. 
The zeroth iteration corresponds to the initial model trained using only the train set; subsequent iterations pertain to models training using
pseudo-labelling. In Fig.~\ref{fig4}, we also show the maximum improvement obtained during the 50 iterations of pseudo-labelling.

The behaviour of the different learning algorithms varies substantially when applying pseudo-labelling. 
\texttt{CatBoostRegressor}, \texttt{LightGBMRegressor}, and \texttt{XGBoostRegressor} all experienced significant 
improvements in model quality. Conversely, pseudo-labelling only offered a marginal improvement to the 
\texttt{KNeighborsRegressor} models, and appears to be significantly detrimental to the \texttt{RandomForestRegressor} 
models.

The reasons why the different learning algorithms respond differently to pseudo-labelling are complex, but can be
 thought of as to a trade-off between the enlargement of the training sample, which usually improves model quality, 
 and the inclusion of some incorrect (pseudo) labels, which can degrade model quality \citep[e.g.,][]{Humphrey2022}.
 A more detailed description of the results is given in the following subsections.

\subsection{Photometric redshift}
\label{photoz}

\subsubsection{Gradient boosted tree models}
\label{gbt_models}
It is clear from Fig.~\ref{fig1} that pseudo-labelling can significantly improve the NMAD and outlier fraction metrics for photometric 
redshift estimates, but this depends on the learning algorithm used, and on the number of iterations of pseudo-labelling that are performed. 

The greatest improvements ($\ga5\%$) are obtained when using a gradient boosted tree method (GBT; \texttt{CatBoost}, \texttt{LightGBM}, or \texttt{XGBoost}; 
see the first 3 rows of Figs.~\ref{fig1}-\ref{fig3}). 
In these cases, the NMAD decreases rapidly in the first $\sim3$ iterations of pseudo-labelling, after which the NMAD curve either flattens-off (\texttt{CatBoost}), 
experiences an up-turn (\texttt{LightGBM}), or continues in a gradual decline (\texttt{XGBoost}). All three GBT algorithms show a broadly similar evolution 
in the catastrophic outlier fraction, with an intially steep decline that becomes shallower as the pseudo-labelling iterations proceed. 

A noteworthy result is that the improvements in the NMAD and outlier fraction of the test set predictions are accompanied by improvements in the 
hold-out set predictions (see, e.g., Fig.~\ref{fig1} panels a,b), indicating that improved generalisation was achieved not only for the test set (used for the 
pseudo-labelling), but also to the hold-out set (not involved in model training). However, the improvements in NMAD and outlier fraction are smaller 
for the hold-out set than for the test set. 

The initial models (zeroth iteration) result in predictions with a generally low redshift bias, which the pseudo-labelling process further improves in the cases of 
\texttt{CatBoost} and \texttt{XGBoost}. 

\subsubsection{Other models}
The results from using pseudo-labelling with the other learning algorithms (\texttt{KNeighborsRegressor} or \texttt{RandomForestRegressor})
are less clear-cut. In the case of \texttt{KNeighborsRegressor} (Fig.~\ref{fig1} panels d,j,p), the first $\sim$2 iterations provide only a marginal improvement in 
NMAD, while the outlier fraction shows a sharp initial decrease followed by a reversal back up towards, but not reaching, the starting value. A marginal improvement 
is also present in the bias. 

The \texttt{RandomForestRegressor} (Fig.~\ref{fig1} panels f,l,r) stands out as the learning algorithm that did not receive any benefit from pseudo-labelling. 
Even from the first iteration of pseudo-labelling, the values of the NMAD, outlier fraction, and bias are always significantly higher than when pseudo-labelling 
is not used (the zeroth iteration). In effect, we see a gradual accumulation of errors due to the introduction of label noise.

\subsection{Stellar mass and SFR}
\label{mass_sfr}
The results from applying pseudo-labelling to the estimation of $M_*$ (Fig.~\ref{fig2}) and SFR (Fig.~\ref{fig3}) are broadly analogous to those 
described in $\S$\ref{photoz}. For the GBT methods, there is a rapid decrease in mean absolute error and median absolute error values during the initial $\sim5$
iterations of pseudo-labelling. The subsequent evolution in these two metrics differs between the three GBT methods. In the case of \texttt{CatBoost}, a constant
value is quickly reached. For (\texttt{LightGBM}), there is a reversal and increase in the metric values. In the case of \texttt{XGBoost}, the metric values continue 
their gradual decrease, albeit with a shallower gradient. Likewise, the $R^2$ scores when using the GBT methods show an initially significant increase in model 
quality (higher values), followed by an evolution that depends on which learning algorithm was used. 

For \texttt{KNeighborsRegressor} the test set predictions improve marginally after a single iteration of the pseudo-labelling procedure, as shown by the decrease in 
the mean absolute error, median absolute error, and $R^2$ score. However, from the second iteration onward model quality dimishes, such that by the tenth iteration the 
aforementioned metrics have values that are significantly worse than the initial (zeroth iteration) model. For the hold-out set, we find little, if any, improvement 
in model quality with the application of pseudo-labelling.

Similar to the results for photometric redshift estimation, in the case of \texttt{RandomForest} the mean absolute error, median absolute error, and $R^2$ score show
a substantial reduction in model quality as the result of application of pseudo-labelling.

\section{Concluding remarks}
\label{conclusions}
We have explored the application of the semi-supervised pseudo-labelling methodology to the problem of estimating galaxy redshift, stellar mass, and 
star-formation rate from broad band photometry and colours. The dataset used is derived from the COSMOS2015 multiwavelength catalogue of \citet{Laigle2016}.
Our tests simulate the case of a photometric dataset produced by a large-scale imaging survey, for which labels are available only for a 
small subset of the sources; the objective is to examine whether pseudo-labelling can be used to improve machine learning models for the prediction of the 
missing labels, compared to the simple case where only the labelled data is used for model training.

We have demonstrated that our methodology can result in regression models that are able to predict redshift, stellar mass, and star-formation rate 
significantly more accurately, with typical improvements in metrics of absolute error of $\sim$5--15\%, by making use of information present in unlabelled data. 

In cases where improved predictions are obtained for the test data used for the pseudo-labelling, the predictions for hold-out data (not used for pseudo-labelling)
also show significant, but typically smaller, improvement. In other words, pseudo-labelling also results in improved generalisation to previously unseen hold-out 
data. 
 
An important caveat that must be stated is that the performance of our implementation of pseudo-labelling, in terms of how well it improves supervised
machine learning models for galaxy properties, depends on the learning algorithm with which it is used. For instance, we find that models produced with 
the gradient boosting tree methods \texttt{CatBoostRegressor}, \texttt{LightGBMRegressor}, and \texttt{XGBoostRegressor} experience the greatest improvements
from the use of pseudo-labelling. Conversely, models trained with the popular tree-ensemble method \texttt{RandomForestRegressor} were negatively affected 
by the application pseudo-labelling. 

This study is intended to be a proof-of-concept for the use of pseudo-labelling for the estimation of the properties of astronomical sources. As such, we 
have not exhaustively tested all available learning algorithms, nor have we exhaustively optimised the hyperparameters of the learning algorithms we use herein. 
In a future study, we will explore in greater detail the application of pseudo-labelling (and other variants thereof) to the classification and property 
estimation of astronomical sources. 

Finally, we emphasize that the pseudo labelling technique is likely to be useful for enhancing the quality of estimates of galaxy redshift and physical properties 
in large upcoming (and ongoing) imaging surveys such as UNIONS \citep{Chambers2020}, LSST \citep{Ivezic2019} and {\it Euclid} \citep{Laureijs2011}, which will 
produce rich, but challengingly vast datasets in need of computationally efficient labelling. 

\section*{Acknowledgements}
We thank the referee for their careful reviews and useful comments. 
This work was supported by Funda\c{c}\~ao para a Ci\^encia e a Tecnologia (FCT) through grants UID/FIS/04434/2019, UIDB/04434/2020, 
UIDP/04434/2020 and PTDC/FIS-AST/29245/2017, EXPL/FIS-AST/1085/2021, and an FCT-CAPES Transnational Cooperation Project. 
R.C. acknowledges support from the FCT through the Fellowship PD/BD/150455/2019 
(PhD:SPACE Doctoral Network PD/00040/2012). We also acknowledge support from NVIDIA in the form of a GPU under the NVIDIA Academic Hardware 
Grant Program. In the development of this work, we have made use of the \texttt{Pandas} \citep{McKinney2010}, \texttt{Numpy} \citep{Harris2020}, and
\texttt{Dask} \citep{Rocklin2015} packages for \texttt{Python}.

\section*{Data Availability}

All data and methods used in this study are publicly available at \href{https://github.com/humphrey-and-the-machine}{https://github.com/humphrey-and-the-machine}.



\bsp	
\label{lastpage}
\end{document}